\documentclass[9pt,twocolumn,twoside]{pnas-new}
\templatetype{pnasresearcharticle}
\usepackage{longtable,array,calc,tabularx,makecell,mathtools,adjustbox}
\usepackage{needspace,xurl}

\providecommand{\real}[1]{#1}

\DeclareUnicodeCharacter{200B}{}
\DeclareUnicodeCharacter{2265}{\ensuremath{\geq}}
\DeclareUnicodeCharacter{2264}{\ensuremath{\leq}}
\DeclareUnicodeCharacter{03BB}{\ensuremath{\lambda}}
\DeclareUnicodeCharacter{03C6}{\ensuremath{\varphi}}
\DeclareUnicodeCharacter{03B8}{\ensuremath{\theta}}
\DeclareUnicodeCharacter{03BA}{\ensuremath{\kappa}}
\DeclareUnicodeCharacter{03C8}{\ensuremath{\psi}}
\DeclareUnicodeCharacter{03B1}{\ensuremath{\alpha}}
\DeclareUnicodeCharacter{03B3}{\ensuremath{\gamma}}
\DeclareUnicodeCharacter{03C1}{\ensuremath{\rho}}
\DeclareUnicodeCharacter{03B4}{\ensuremath{\delta}}
\DeclareUnicodeCharacter{2212}{\ensuremath{-}}
\DeclareUnicodeCharacter{00D7}{\ensuremath{\times}}
\DeclareUnicodeCharacter{221E}{\ensuremath{\infty}}
\DeclareUnicodeCharacter{03C4}{\ensuremath{\tau}}
\DeclareUnicodeCharacter{03B7}{\ensuremath{\eta}}
\DeclareUnicodeCharacter{03BC}{\ensuremath{\mu}}
\DeclareUnicodeCharacter{03C3}{\ensuremath{\sigma}}
\DeclareUnicodeCharacter{03BD}{\ensuremath{\nu}}
\DeclareUnicodeCharacter{03BE}{\ensuremath{\xi}}
\DeclareUnicodeCharacter{03B5}{\ensuremath{\epsilon}}
\DeclareUnicodeCharacter{03B6}{\ensuremath{\zeta}}
\DeclareUnicodeCharacter{03C0}{\ensuremath{\pi}}
\DeclareUnicodeCharacter{2113}{\ensuremath{\ell}}
\DeclareUnicodeCharacter{2032}{\ensuremath{\prime}}
\DeclareUnicodeCharacter{00B3}{\textsuperscript{3}}
\DeclareUnicodeCharacter{25AB}{\ensuremath{\square}}
\DeclareUnicodeCharacter{207B}{\textsuperscript{-}}
\DeclareUnicodeCharacter{2075}{\textsuperscript{5}}
\DeclareUnicodeCharacter{00B9}{\textsuperscript{1}}
\DeclareUnicodeCharacter{00B2}{\textsuperscript{2}}
\DeclareUnicodeCharacter{1D62}{\textsubscript{i}}
\DeclareUnicodeCharacter{2080}{\textsubscript{0}}
\DeclareUnicodeCharacter{208A}{\textsubscript{+}}
\DeclareUnicodeCharacter{221A}{\ensuremath{\surd}}
\setboolean{displaywatermark}{false}
\fancypagestyle{cwe}{\fancyhf{}\fancyfoot[L]{\footerfont Falco}\fancyfoot[R]{\footerfont Preprint\hspace{7pt}|\hspace{7pt}\thepage\ /\ \pageref{LastPage}}}
\begin{document}
\title{How a group that cannot communicate knows itself by its enemy}
\author[a,1]{Gregory Falco}
\affil[a]{Sibley School of Mechanical and Aerospace Engineering, Cornell University, Ithaca, NY 14853}
\leadauthor{Falco}
\significancestatement{When hunters cannot talk, each may learn the other's position from the animal they pursue. The animal need not help them willingly: hiding their situation can interfere with its own escape. We show how a team can use this conflict by agreeing beforehand on how to respond. A sufficiently cautious response can make the opponent reveal the truth, even when the team cannot safely use all of it. This makes the opponent's reaction an enemy-made counterpart of a price, conveying knowledge that no member holds alone. Differences in how members respond can replace private randomization, while a small discount on trust provides a quantified safeguard against errors about the opponent.}
\authorcontributions{Author contributions: G.F. conceived and designed the research, developed the theoretical framework (the research question, model and definitions), directed the analysis, interpreted the analytical and computational results, and wrote the paper.}
\authordeclaration{The author declares no competing interest.}
\correspondingauthor{\textsuperscript{1}To whom correspondence should be addressed. E-mail: gfalco@cornell.edu}
\begin{abstract}
Teams that cannot communicate seem stuck with what each member sees. We show that their opponent can tell them the rest. Silent hunters, jammed robots and traders who never meet all face an adversary that reacts to the group as a whole; every member can read that reaction, a tell, and learn what no member knows alone. One tell can do the work of a message from every member. The adversary would rather hide, but hiding interferes with its own task, and a team that commits in advance to how it will respond decides how much hiding is worth. A committed team never loses when its opponent cares more about its own goals, and does as well as a fully informed team exactly when the opponent's stake outweighs any gain from lying. Below that point the team learns more by trusting less: in a solved Gaussian model it obtains the truth by promising not to use all of it and keeps the cost the opponent would have paid to conceal; an exact test decides when this holds in finite games. How a team spreads its trust matters too. A lone uninformed member with a full-support prior needs a private coin, but with additive squared-error losses two interchangeable members who respond differently replace it, and a small discount on trust insures against misjudging the enemy. Information, on this view, is not only gathered or aggregated; it is produced by how a team commits to respond.
\end{abstract}
\dates{Preprint, October 2026. SI Appendix available as an ancillary file. Code and data: doi:10.5281/zenodo.23743441}
\doi{}
\maketitle
\thispagestyle{cwe}
\ifthenelse{\boolean{shortarticle}}{\ifthenelse{\boolean{singlecolumn}}{\abscontentformatted}{\abscontent}}{}
\Firstpage
\setlength{\abovedisplayskip}{6pt plus 2pt minus 2pt}
\setlength{\belowdisplayskip}{6pt plus 2pt minus 2pt}
\setlength{\belowdisplayshortskip}{4pt plus 1pt minus 1pt}
\setlength{\dbltextfloatsep}{12pt plus 2pt minus 2pt}

Sun Tzu wrote, ``If you know the enemy and know yourself, you need not fear the result of a hundred battles'' (1). This paper formalizes a reversal of that maxim: a silent army can know itself by its enemy. An opponent's reaction can reveal facts about the group that none of its members knows alone.

Imagine two hunters who cannot talk, each seeing only part of the situation while the animal they pursue can see both. As the animal moves, each hunter may learn something about the other's position. But the animal has a reason to conceal that information if it expects the hunters to exploit it aggressively. Suppose instead that the hunters agree in advance to respond more cautiously, and the animal knows their rule. Deception then becomes less rewarding, so a revealing reaction may serve the animal better. How the hunters promise to react can change what they get to see.

The animal need not cooperate for this to work. Because it must pursue its own escape, hiding the hunters' situation can come at a cost to its own task. By choosing their responses before the animal moves, the hunters can change the balance between that cost and the benefit of misleading them.

We establish when this conflict makes the opponent a useful relay for the team's scattered information, and when it allows the team to perform as well as if fully informed. The surprising possibility is that a team can obtain the truth by committing to react less strongly to it.

Hayek described prices as a way to use knowledge ``not given to anyone in its totality'' (2). Sun Tzu identified the army's problem: ``the spoken word does not carry far enough: hence the institution of gongs and drums'' (1). A price is designed by no one and a drum by one's own commander. This paper concerns a third public signal, made by the enemy. Prey reacts to a pack, a dealer to combined orders, and an intruder to a network's defenses. Each reaction can reveal a fact that no member knows alone. We call the reacting party a \emph{witness} and its observable reaction a \emph{tell}. This enemy-made counterpart of a price turns on a simple comparison: a lie gains the witness \(\psi L\) in harm to the team but costs it \(\kappa d\) in its own aims.

What a team gets to observe is therefore part of the strategic problem. When the team changes its promised responses, it changes the witness's incentive to reveal or conceal, and hence the information that reaches its members. Commitment can shape a common observation even though the members never send one another a message.

\parabreak

The distinctive problem arises when members' observations jointly determine a fact that none knows alone. The witness's public reaction can relay that fact among separated decision makers (3). We measure this relay and compare it with member broadcasts, then establish adversity efficiency and the threshold for full-information performance. Below that threshold, the witness paradox separates revealing the truth from acting fully on it. The exact Gaussian solution shows how differences in members' responses can replace private noise, while a discount on trust provides a quantified margin of safety.

The analysis connects price aggregation (2, 4, 5, 6, 7, 8, 9) and revelation through actions (10, 11, 12, 13, 14) with team theory (15, 16, 17, 18, 19), equilibrium and common knowledge (20, 21, 22), commitment (23, 24), and information theory (25, 26, 27). Whereas silent teams can coordinate through an outside signal they learn to use (28), here the opponent chooses the signal in response to their policies.

\begin{table*}[t]
\centering
\caption{What each feature of the setting contributes.}
\label{tab:restrictions}
\begin{tabular}{@{}
  >{\raggedright\arraybackslash}p{(\textwidth - 6\tabcolsep) * \real{0.2500}}
  >{\raggedright\arraybackslash}p{(\textwidth - 6\tabcolsep) * \real{0.2500}}
  >{\raggedright\arraybackslash}p{(\textwidth - 6\tabcolsep) * \real{0.2500}}
  >{\raggedright\arraybackslash}p{(\textwidth - 6\tabcolsep) * \real{0.2500}}@{}}
\toprule\noalign{}
\begin{minipage}[b]{\linewidth}\raggedright
\textbf{If the setting changes this way}
\end{minipage} & \begin{minipage}[b]{\linewidth}\raggedright
\textbf{Closest existing theory}
\end{minipage} & \begin{minipage}[b]{\linewidth}\raggedright
\textbf{What still holds}
\end{minipage} & \begin{minipage}[b]{\linewidth}\raggedright
\textbf{What is lost}
\end{minipage} \\
\midrule\noalign{}

Members pool observations that determine \(\varphi\), or talk & One decision maker; costly persuasion (36, 37) & The witness's own task & Relay, the threshold, the paradox, diversity \\
Members share a coin & Teams with a shared coin facing an opponent (17, 18) & Every commitment value when losses add up & Gaps in existence when both sides move at once \\
The witness's information is independent of members' joint observations & Signaling and persuasion (11, 36) & Theorems 2 to 4 & Relay among members; replacing broadcasts \\
Only one member & Receiver commitment and randomization (33, 34, 35) & The exact Gaussian value; the cost of deterministic rules & Relay; diversity replacing noise \\
A designer, not the opponent, chooses the signal & Outside coordination signals (28) & The value of a shared signal & Adversity efficiency, the threshold, the paradox \\
\bottomrule
\end{tabular}
\end{table*}

\subsection*{The witness game}

\textbf{Definition 1 (Witness game).} A team \(\mathcal{N} = \{ 1,\ldots,N\}\) faces a witness in a state \(s \sim P\) drawn by nature. Member \(i\) observes only its own piece of information, \(o_{i}(s)\), whereas the witness observes \(\varphi = \varphi(s)\); if it sees more, this variable denotes all its information. The witness chooses a public action \(w \in W\) according to a strategy \(\sigma( \cdot \mid \varphi)\). Each member then observes a tell \(y \sim \mathcal{K}( \cdot \mid w)\), with \(y = w\) as the default case. Conditional on \(w\), the tell is independent of the members' private observations, and member \(i\) chooses its action using \(\pi_{i}\left( a_{i} \mid o_{i},y \right)\). The members share a payoff \(u_{T}(a,s)\), while the witness's payoff is (Fig. 1\emph{A})

\[u_{W} = \kappa\, g(w,a,s) - \psi\, u_{T}(a,s) + c(\varphi,s).\]

The witness weights its own objective \(g\) by \(\kappa > 0\) and harm to the team by \(\psi \geq 0\), while \(c\) is choice-independent. Their balance is described by \(\theta = \kappa/\psi\). When \(g = u^{0}(w,\varphi)\), the favorite action is \(b(\varphi) = \arg\max_{w}u^{0}(w,\varphi)\) and the regret \(d(w,\varphi) = u^{0}\left( b(\varphi),\varphi \right) - u^{0}(w,\varphi)\) is the cost of choosing \(w\) instead. Full information gives the team \(V^{*} = \mathbb{E}\max_{a}u_{T}(a,s)\).

We assume the witness observes the relevant group fact, even when no member can determine it alone, and study its transmission rather than its acquisition.

\textbf{Assumption 1 (Silence).} Members cannot communicate or use a shared randomization device during play, although each may randomize privately. They may agree on their policies beforehand, and those policies are known to the witness. If members could instead communicate without restriction, they would pool their observations into a single decision maker, which knows \(\varphi\) when those observations jointly determine it (\emph{SI Appendix}, section S6).

\textbf{Definition 2 (Common witness equilibrium).} The team commits to its members' response policies before the witness acts, anticipating that the witness will choose a best response. Let \(\Pi\) contain the policy profiles in which each member uses only its own observation, the common tell, and its own randomization. Write \(BR(\pi)\) for the witness's best responses to \(\pi\). A common witness equilibrium (CWE) is a pair \(\left( \pi^{*},\sigma^{*} \right)\) with \(\sigma^{*} \in BR\left( \pi^{*} \right)\) that attains

\[U_{T}\left( \pi^{*},\sigma^{*} \right) = \max_{\pi \in \Pi}\ \max_{\sigma \in BR(\pi)}U_{T}(\pi,\sigma).\]

This definition selects ties in the team's favor. A CWE is \emph{robust} when the witness has a unique best response, so its attainment does not rely on that selection. For comparison, a \emph{reflex} witness follows a strategy that is unaffected by the team's choice of policies.

\textbf{Theorem 1 (Existence).} When information and member action sets are finite and the witness's action set is compact with continuous payoffs, a CWE exists. Its value is at least that of every simultaneous equilibrium with independently randomized member policies (\emph{SI Appendix}, section S2).

\subsection*{Information relayed among separated members}

Proposition 1 (Adversarial relay). (a) Relay. Suppose \(\varphi\) is determined by member \(i\)'s observation together with those of a group \(J\), and the tell depends on the state only through \(\varphi\). Every witness strategy with expected task regret at most \(D\) then relays information between \(i\) and \(J\) according to the bound

\[I\left( o_{i};y \mid o_{J} \right)\  \geq \ I(\varphi;y) - I\left( \varphi;o_{J} \right)\  \geq \ R^{obs}(D) - I\left( \varphi;o_{J} \right),\]

Here information is in nats, and \(R^{obs}(D)\) is the least information about \(\varphi\) an admissible tell can carry at expected regret at most \(D\). The bound measures what stealth and deception cannot hide; a payoff gain additionally requires the team to use the information effectively.

(b) Provocation. If the witness ignores the team's earlier actions, and those actions change neither its information nor the team's final payoff, costly probing cannot help. In this case there is no strategic reaction for the probe to reveal.

(c) Broadcast replacement. In the symmetric Gaussian team with \(o_i=s_i\) and \(\varphi=\sum_j s_j\), consider a jointly Gaussian tell carrying \(0<I<\infty\) nats. If \(R^{*}\) is the least total information rate of one-round member broadcasts using private noise that attain the same team value, then \(N - 1 \leq R^{*}/I \leq N\). The ratio is exactly \(N\) for independent fair bits with an exactly revealing parity tell. These comparisons concern the stated one-round broadcast benchmarks; they do not assert the same rates for arbitrary interactive protocols (proofs in \emph{SI Appendix}, section S6).

In the parity example, each member knows one independent coin flip, but everyone needs to know whether the total number of heads is odd. One public bit supplies the answer that one-round broadcasts require \(N\) bits to convey. The witness's task constraint can force it to supply that bit unwillingly.

Pooling observations would make the relay unnecessary (\emph{SI Appendix}, Proposition S11a). Unlike signaling through teammates' actions (29, 30), the tell comes from someone who may want the team ignorant. Sun Tzu's ``Force him to reveal himself'' (1) describes its active version: probing changes the reaction the team reads (31, 32).

\subsection*{What existing theory already explains}

Receiver commitment already extracts information by binding a decision maker to responses it would not choose afterward (33), and its connection to randomization is known for one sender and receiver (34, 35). Our single-member results sharpen that logic, while the team results address relay among separated members. Table 1 identifies which conclusions depend on separation, a collective state, or an opponent-made signal (\emph{SI Appendix}, Proposition S11).

\section*{Results}

\subsection*{Adversity efficiency}

The first result asks what happens when the opponent puts more weight on its own task while the team's response is held fixed. Let \(\varphi\) denote everything the witness observes, including any earlier team actions, and let its objective \(g\) be any function of \((w,a,s)\). The stake ratio \(\theta(\varphi)\) may depend on \(\varphi\). The team commits to \(\pi\) from a fixed feasible set \(\Pi\), which may contain policies that respond over time or lotteries over plans. Define the conditional payoffs \(G_{\pi}(w,\varphi) = \mathbb{E}\lbrack g \mid \varphi,w,\pi\rbrack\) and \(T_{\pi}(w,\varphi) = \mathbb{E}\left\lbrack u_{T} \mid \varphi,w,\pi \right\rbrack\). The witness then chooses from \(BR_{\theta}(\pi,\varphi) = \arg\max_{w}\left\lbrack \theta(\varphi)G_{\pi}(w,\varphi) - T_{\pi}(w,\varphi) \right\rbrack\), which we assume is nonempty.

Theorem 2 (Adversity efficiency). Let \(\theta \leq \theta\prime\) pointwise. For any committed policy \(\pi\) and any witness observation \(\varphi\) at which \(\theta(\varphi) < \theta\prime(\varphi)\), all best responses \(\sigma \in {BR}_{\theta}(\pi,\varphi)\) and \(\sigma\prime \in {BR}_{\theta\prime}(\pi,\varphi)\) satisfy

\[\begin{aligned}
G_{\pi}(\sigma',\varphi)&\geq G_{\pi}(\sigma,\varphi),\\
T_{\pi}(\sigma',\varphi)&\geq T_{\pi}(\sigma,\varphi).
\end{aligned}\]

Thus an increase in the stake ratio cannot lower the team's committed value, regardless of how best-response ties are resolved, with the same selection at unchanged ratios.

\emph{Proof.} Hold \(\pi\) and \(\varphi\) fixed and suppress them in the notation. Since \(\sigma\) is optimal at \(\theta\) and \(\sigma\prime\) is optimal at \(\theta\prime\), we have

\[\begin{aligned}
\theta G(\sigma)-T(\sigma)&\geq\theta G(\sigma')-T(\sigma'),\\
\theta' G(\sigma')-T(\sigma')&\geq\theta' G(\sigma)-T(\sigma).
\end{aligned}\]

Adding the inequalities yields \((\theta\prime - \theta)\,\left\lbrack G(\sigma\prime) - G(\sigma) \right\rbrack \geq 0\), which implies \(G(\sigma\prime) \geq G(\sigma)\). Substituting this ordering into the first inequality gives \(T(\sigma\prime) - T(\sigma) \geq \theta\,\left\lbrack G(\sigma\prime) - G(\sigma) \right\rbrack \geq 0\). At observations where \(\theta(\varphi) = \theta\prime(\varphi)\), the best-response problem is unchanged and a fixed selection gives the same response. Taking expectations over \(\varphi\) and then optimizing over \(\Pi\) establishes the value comparison. \(▫\)

An opponent with more at stake in its own task is less willing to sacrifice it merely to hurt the team. This orders team payoff, but neither informativeness nor agreement among members must improve; \emph{SI Appendix}, section S3 and Table S9, give the counterexamples.

\subsection*{The witness first welfare theorem}

To characterize full-information performance, set \(y = w\) and \(g = u^{0}(w,\varphi)\). Under (A1), the unique optimum can be implemented using only \(\varphi\) and members' private observations, as \(a^{*}(s) = \left( a_{i}^{*}\left( \varphi(s),o_{i} \right) \right)_{i}\). Under (A2), the witness's favorite action is unique and differs across \(\varphi\). Under (A3), the tell depends only on \(w\) and witness randomization is independent of members' observations given \(\varphi\). Condition (D) permits deterrence of actions outside \(b(\Phi)\). The converse additionally requires (A4): every private observation is possible under every \(\varphi\).

Let \(L(\varphi\prime \mid \varphi)\) be the team's loss when it acts as if the witness had type \(\varphi\prime\) although its true type is \(\varphi\). The impersonation ratio \(\lambda^{*}\) compares the harm from that lie with the witness's own cost:

\[\begin{aligned}
L(\varphi'\mid\varphi)&=\mathbb E\bigl[u_T(a^*(s),s)\\
&\quad-u_T(a^*(\varphi',o),s)\mid\varphi\bigr],\\
\lambda^*&=\sup_{\varphi\ne\varphi'}\frac{L(\varphi'\mid\varphi)}{d(b(\varphi'),\varphi)}.
\end{aligned}\]

Theorem 3 (Witness first welfare theorem). Assume (A1)--(A3) and (D). If \(\theta \geq \lambda^{*}\), an honest witness choosing \(w = b(\varphi)\) and members following \(a_{i}^{*}\left( b^{- 1}(w),o_{i} \right)\) form a CWE with value \(V^{*}\). The same outcome is a simultaneous equilibrium, and full revelation is optimal when the witness commits to its channel. These outcomes are robust if \(\theta > \lambda^{*}\) and (D) holds strictly. Conversely, under (A4), any outcome in which the witness best-responds and the team earns \(V^{*}\) must have an honest witness and \(\theta \geq \lambda^{*}\). Consequently, \(V_{CWE} = V^{*}\) if and only if \(\theta \geq \lambda^{*}\).

\emph{Proof idea.} At value \(V^{*}\), the witness's hostility term is at its minimum. Moving to its favorite action cannot lower that term, so any dishonest action admits a profitable deviation. Honesty requires every impersonation to be unprofitable, giving \(\theta \geq \lambda^{*}\). Sufficiency verifies these incentives and off-range deterrence (\emph{SI Appendix}, section S4). \(▫\)

\textbf{Corollary 1 (Gaussian team).} Consider \(\varphi \sim \mathcal{N}(0,1/\alpha)\), with precision \(\alpha\) (one over the variance). Members receive Gaussian private signals \(x_{i} = \varphi + \varepsilon_{i}\) with precision \(\gamma\), the witness's own objective is \(u^{0} = - \frac{1}{2}(w - k\varphi)^{2}\), and the team's payoff is \(u_{T} = - \sum_{i}^{}\left( a_{i} - \varphi \right)^{2} - \eta\sum_{i < j}^{}\left( a_{i} - a_{j} \right)^{2}\). A team fooled into accepting \(\varphi\prime\) acts in unison on that report, giving the impersonation ratio \(\lambda^{*} = 2N/k^{2}\). The team reaches \(V^{*}\) under team commitment, witness commitment, or simultaneous equilibrium if and only if

\[\kappa k^{2}\  \geq \ 2\psi N.\]

The threshold compares the witness's own stake with the harm it can impose on the team. Prior uncertainty, signal precision, coordination and dispersion matter only below the threshold. The model covers dispersed information exactly: if member \(i\) holds \(s_{i}\), where the \(s_{j}\) are independent, identically distributed, centered normal components of \(\varphi = \sum_{j}s_{j}\), then \(x_{i} = Ns_{i}\) is a signal of \(\varphi\) with precision \(\alpha/(N - 1)\), and with \(\eta = 0\) every Gaussian result applies (\emph{SI Appendix}, Corollary S6).

The condition is the familiar no-mimicking constraint (38), as with lying costs (39, 40), cheap talk (41) and verifiable disclosure (42, 43); witness commitment is costly persuasion (36, 37), team commitment a receiver-chosen rule (35).

\subsection*{The witness paradox}

Falling below the threshold rules out full-information performance under the theorem's assumptions, but it need not rule out a fully revealing tell. The team may instead obtain the truth by committing not to act on all of it. To make this distinction exact, consider the Gaussian team with \(\eta = 0\) (\emph{SI Appendix}, Theorem S7 treats \(\eta > 0\)). Write \(\rho = 2\psi N/\left( \kappa k^{2} \right) = \lambda^{*}/\theta\), so that efficiency requires \(\rho \leq 1\), and define \(M = \sqrt{\rho}/\left( \sqrt{\rho} - 1 \right)\) for the below-threshold regimes.

Theorem 4 (Witness paradox). Let \(\theta < \lambda^{*}\).

\begin{enumerate}
\def\labelenumi{(\alph{enumi})}
\item
  \emph{Both move at once.} Under (A1)--(A4), no team-optimal equilibrium has a fully revealing witness. In the Gaussian team, a linear Nash equilibrium with the outcome in (c) exists for every \(\rho > 1\).
\item
  \emph{The team commits.} The exact Gaussian loss per member is \(\ell^{*} = \max_{0 \leq u \leq 1/\alpha}\left\lbrack u/(1 + \gamma u) - u/\rho \right\rbrack\). It equals \(1/\left( M^{2}\gamma \right)\) when \(\sqrt{\rho} \leq 1 + \gamma/\alpha\) and \(1/(\alpha + \gamma) - 1/(\rho\alpha)\) otherwise. This is the witness-commitment loss minus \(\theta\) times the witness's task regret at its own commitment optimum. In the interior of the first regime, the unique optimal rule is committed skepticism: \(a_{i} = (1 - c)x_{i} + c\, w/k\) with trust \(c = 1/\sqrt{\rho}\). In the second regime, members retain the private-signal weight they would use without a tell, assign the tell weight \((\alpha + \gamma)/(\rho\alpha)\), and add private noise or, for \(N \geq 2\), distribute trust unequally across members. Exact attainment selects honesty from an indifferent witness's best responses. Trust slightly below \(1/\sqrt{\rho}\) instead makes honesty strict.
\item
  \emph{The witness commits.} The optimal Gaussian channel shrinks and adds noise to the report. It reveals everything if \(\rho \leq 1\), conceals everything if \(\sqrt{\rho} \geq 1 + \gamma/\alpha\), and otherwise leaves the team with precision \(M\gamma\) (\emph{SI Appendix}, Lemma S3 and section S5).
\end{enumerate}

With optimal team commitment the witness tells the truth, but when \(\sqrt{\rho} \leq 1 + \gamma/\alpha\), members give that report less weight than they would if free to revise their responses afterward. The resulting effective precision is \(M^{2}\gamma\), compared with \(M\gamma\) under witness commitment (Fig. 1B). Beyond this regime, honesty also requires private noise or different trust weights.

\begin{figure*}[t]
\centering
\includegraphics[width=\textwidth]{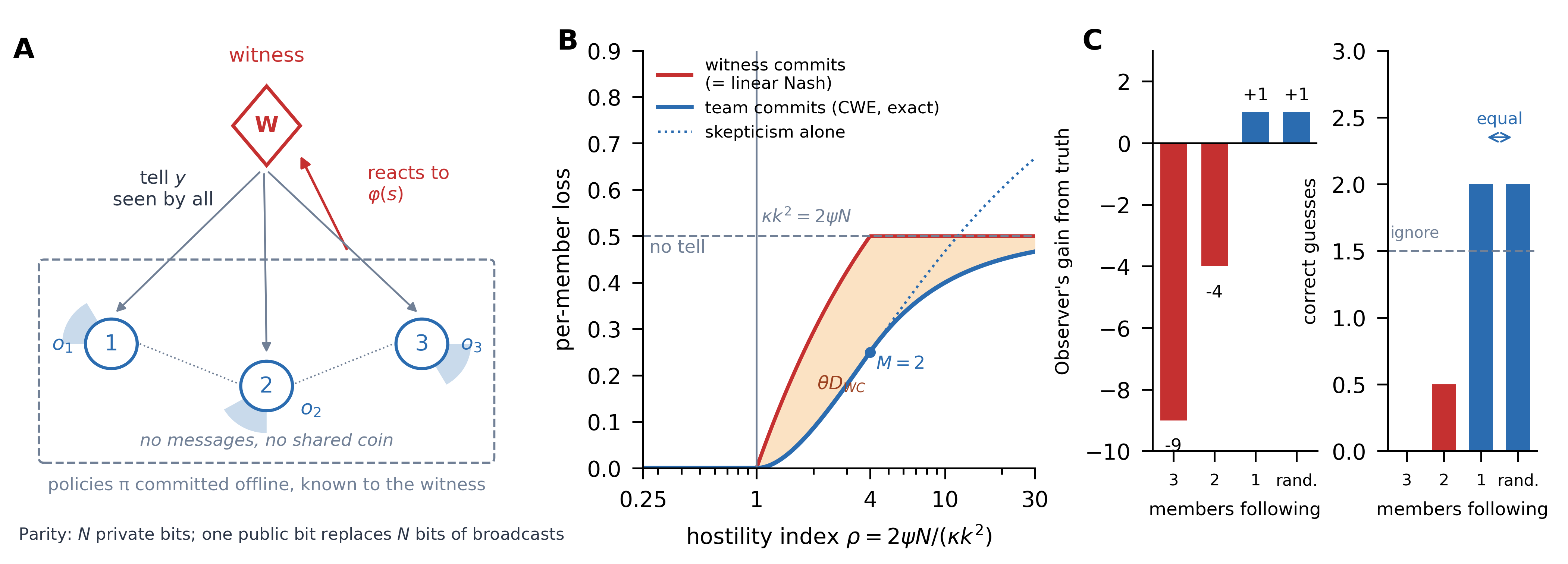}
\caption{Adversarial relay, commitment and the diversity test. (A) Members, each with private information \(o_{i}\), cannot talk; their rules, fixed in advance, are known to the witness, which reacts to the group fact \(\varphi(s)\) with a tell \(y\) that every member sees. (B) Loss per member in the Gaussian team (\(\alpha = \gamma = 1\)) as hostility \(\rho = 2\psi N/\left( \kappa k^{2} \right)\) rises: red shows witness commitment, also attained in linear Nash play; blue shows the exact team-commitment loss, and the dotted curve shows skepticism alone. The shaded band is the hiding cost \(\theta D_{WC}\) that commitment lets the team keep. (C) Predicted results of the diversity test (\(\theta = 1.2\)) by the number of Members committed to follow the report (rand., each follows with probability 1/3).}
\label{fig:relay}
\end{figure*}

The value of commitment has a precise interpretation in the Gaussian model:

\[V^{*} - V_{CWE}\mspace{6mu} = \mspace{6mu}\left( V^{*} - V_{WC} \right) - \theta\, D_{WC}.\]

Here \(D_{WC}\) is the witness's task regret when it commits to conceal. Team commitment recovers exactly that cost, weighted by the stake ratio.

This reverses the incentive problem in Grossman and Stiglitz (44), where fully revealing prices remove the incentive to acquire information. Here, fully using a revealing tell destroys its producer's incentive to reveal. Commitment restores that incentive by limiting the response. The witness may thus hide its audience's collective state, rather than an insider's secret (10) or a government's private information (13).

\subsection*{Noise and diversity}

Private noise can be essential when the team needs deterministic policies. For one uninformed member, \(\rho > 1\) and a full-support prior imply that every deterministic rule remains strictly worse than randomized commitment, even under approximation (\emph{SI Appendix}, Theorem S12 and Proposition S8). With several members, however, differences in their rules can provide the required spread in actions.

\textbf{Theorem 5 (Diversity replaces noise).} Suppose the team's loss is \(\sum_{i}\ell\left( a_{i},\varphi \right)\), where \(\ell\) is quadratic in the action with coefficients that may depend on \(\varphi\). Let hostility be aligned and \(g = u^{0}(w,\varphi)\). Consider a group of \(n \geq 2\) members whose observations are identically distributed conditional on \(\varphi\) and the tell. Suppose each uses the same privately randomized rule, with conditional mean \(m(o,y)\) and variance \(v(o,y)\). Choose weights \(s_{i}\) that sum to zero and whose squares sum to \(n\). If actions are unrestricted real numbers, or if every constructed action \(m(o,y) + s_{i}\sqrt{v(o,y)}\) is feasible, the deterministic profile \(a_{i} = m\left( o_{i},y \right) + s_{i}\sqrt{v\left( o_{i},y \right)}\) preserves every witness type's payoff from every action and preserves the team's value. Any value attained with that private noise is therefore attained without it. For two members, \(s = \pm 1\) gives the paired construction.

\emph{Proof.} A quadratic loss depends on a randomized action only through its first two conditional moments. In the sum across members, the constructed deterministic rules reproduce the variance contribution because their squared weights sum to \(n\). The remaining cross terms cancel because their expectations are common across these members and the weights sum to zero. The team's payoff and every witness type's payoff are therefore unchanged. \(▫\)

In the trust constructions, stronger and weaker responses to the tell reproduce the effect of a private coin. The theorem requires neither Gaussian information nor linear rules; \emph{SI Appendix}, Theorem S13, gives the Gaussian application.

\textbf{Proposition 2 (Trust as many members as the stake is worth).} Let \(\varphi\) be a fair coin, with \(N\) uninformed members each guessing its outcome after a public report. The team's payoff is the number of correct guesses. The witness earns \(\kappa\) for reporting truthfully and loses \(\psi\) for each correct member, giving \(\theta = \kappa/\psi\). Private randomization allows the team to attain \(N/2 + \min\{\theta,N\}/2\). Under deterministic commitment, the best plan instead assigns \(\min\{\lfloor\theta\rfloor,N\}\) members to follow the report while the rest ignore it, attaining \(N/2 + \min\{\lfloor\theta\rfloor,N\}/2\). If all members must use an identical deterministic rule, the value is only \(N/2\) unless \(\theta \geq N\) (\emph{SI Appendix}, section S17).

Diversity buys trust in whole members, whereas randomization buys fractions, so with binary choices diversity cannot always replace a coin. When \(N = 2\) and \(\theta = 2/3\), randomized policies earn \(4/3\) but every deterministic plan earns 1.

\subsection*{Robust commitment}

The exact optimum leaves the witness indifferent, making it vulnerable to errors about the opponent or departures from exact optimization. Trusting the tell slightly less gives honesty a strict advantage, at a performance cost the next theorem quantifies.

\textbf{Theorem 6 (Robust commitment).} Let actions be real. Member \(i\) observes \(x_{i} = \varphi + \varepsilon_{i}\) and reads the tell as \({\widehat{q}}_{i} = w/k + \nu_{i}\). Design for normalized hostility-to-stake ratio \(\bar{\rho}\) and trust discount \(\epsilon \in \lbrack 0,1)\) by using \(a_{i} = (1 - z)x_{i} + c\,{\widehat{q}}_{i} + \xi_{i}\), where \(z \geq 1/\sqrt{\bar{\rho}}\), \(c = (1 - \epsilon)/\left( \bar{\rho}z \right)\), and \(\xi_{i} = \sqrt{zc - c^{2}}\,{\widehat{q}}_{i}\,\zeta_{i}\). Assume the underlying seeds \(\varepsilon_{i}\), \(\nu_{i}\), and \(\zeta_{i}\) are centered, with variances \(\sigma_{\varepsilon}^{2}\), \(s^{2}\), and 1. They are mutually independent, independent across members, and independent of \(\varphi\) and witness randomization. The prior has known mean zero and variance \(\sigma^{2}\), and \(\eta = 0\). If the true ratio is \(\rho\), suppose the witness's expected payoff is within \(\delta\) of its best response, in units of \(\kappa k^{2}/2\). Then (i) whenever \(\rho < \bar{\rho}/(1 - \epsilon)\), reporting \(q \neq \varphi\) rather than truthfully reduces its payoff by \(m(q - \varphi)^{2}\), where \(m = 1 - (1 - \epsilon)\rho/\bar{\rho}\). If instead \(\rho > \bar{\rho}/(1 - \epsilon)\), exaggeration raises its payoff without bound. (ii) In case (i), each member's loss is bounded above by

\[\left( z^{2} - 1/\bar{\rho} \right)\sigma^{2} + (1 - z)^{2}\sigma_{\varepsilon}^{2} + \frac{\epsilon\sigma^{2}}{\bar{\rho}} + \frac{(1 - \epsilon)s^{2}}{\bar{\rho}} + \frac{(1 - \epsilon)\delta}{\bar{\rho}m}.\]

For any prior with the stated moments, the guarantee requires only an upper bound on the normalized hostility-to-stake ratio. A discount \(\epsilon\) buys tolerance for underestimating that ratio by a factor \(1/(1 - \epsilon)\) at cost \(\epsilon\sigma^{2}/\bar{\rho}\) per member; this tolerance is sharp. With matched ratios and a noiseless tell, the loss-bound minimizer is \(\sqrt{\delta/\sigma^{2}}\) when admissible, so caution scales with the square root of the opponent's optimization error (\emph{SI Appendix}, section S17).

The Gaussian equality need not generalize. In finite games, an exact gap certificate identifies policies attaining the concealment bound; with additive losses, a signed-channel dual tests whether an optimum has nonnegative self-weights. A finite counterexample proves a strict gap despite private randomization, while separate non-Gaussian sufficient classes attain the bound (\emph{SI Appendix}, Theorems S14 to S16 and Proposition S10).

\subsection*{A pursuit game}

In the solved pursuit game, each pursuer knows its own randomly chosen side of an evader, but neither knows whether their positions flank it (45). An evader with an open escape route flees under pressure, choosing stealth \(h\) at cost \(\kappa\chi h^{2}/2\). The optimal committed plan has both pursuers close only when no movement is seen, inducing \(h^{*} = \psi\ell/(\kappa\chi)\). Under a hard press the team earns 0.46875, against 0.25 without the tell and 0.5 with communication, and never miscoordinates. The evader's stealth lottery is a correlation source it would rather withhold. Yet its zero marginal cost of stealth at zero means that some concealment remains worthwhile at every finite stake ratio (\emph{SI Appendix}, section S7).

\subsection*{A test of the diversity mechanism}

A proposed experiment isolates diversity from relay. An Observer reports a coin to three Members with announced response plans, earning 6 points for truth and losing 5 per correct Member. Thus \(\theta = 1.2\), and truth is optimal when at most one Member follows. The theory predicts equal truth-telling when one Member always follows and when all three independently follow with probability 1/3, but lying when two or three always follow (Fig. 1C). Honesty preferences (46) should shift the arms in the same direction. Because Members have no private information, the test concerns incentives rather than relay (\emph{SI Appendix}, section S10).

\subsection*{Scope and open questions}

The model requires the witness to observe the relevant group fact and members to observe a common tell; its motivating applications have not been empirically validated here. Full-information efficiency additionally needs (A1)--(A4) and (D), adversity efficiency needs commitment with a fixed own objective and aligned hostility, and exact Gaussian attainment uses favorable ties. Open problems, including the best deterministic rule for one uninformed member and general conditions for tightness, are listed in \emph{SI Appendix}, section S16.

\section*{Discussion}

\emph{For an engineer} designing a silent team of robots or satellites, the opponent becomes a sensor whose output depends on the team's controller. The threshold \(\lambda^{*}\) determines when deception is too costly for full-information coordination. Below it, limiting the controller's response can preserve honesty, with Theorem 6 quantifying a margin against uncertainty about the opponent. Under Theorem 5's conditions, different response rules across agents replace private randomization, so no agent needs to generate noise.

\emph{For an economist}, the information available to an organization becomes an outcome of its commitment. In dynamic moral hazard, a principal may optimally withhold feedback to sustain effort (47); here the problem is reversed: the receivers restrict their own response to sustain an adversary's honesty. Commitment changes the incentive to produce that response, while diversity in members' policies expands what the organization can credibly promise.

An opponent can turn a silent team's dispersed knowledge into a public tell because concealing it interferes with its own task. In the Gaussian model, the team induces honesty and recovers the witness's concealment cost by restraining its use of the information. A committed army can face no concealment at all, only a truth it has committed not to use in full.

\section*{Materials and Methods}

Unless stated otherwise, sets are finite and members share a common prior. Full proofs, model specifications, the script index (S11) and both experimental designs with their oTree implementations (S10) are in the \emph{SI Appendix}. Automated bot tests assess the software; no human-subject results are reported here.

\emph{Use of AI tools.} Under the author's direction, Claude Opus 5.5 (Anthropic) derived and wrote the mathematical proofs, wrote the deposited verification code and edited the manuscript; ChatGPT Sol 6.1 (OpenAI) was used to review and edit the manuscript. The author takes full responsibility for the content.

\emph{Data, Materials, and Software Availability.} All verification scripts, their outputs and the oTree software are deposited in Zenodo (\url{https://doi.org/10.5281/zenodo.23743441}).

\section*{Acknowledgments}

This work was supported by the Andrew Carnegie Foundation (grant G-PS-26-63675).


\end{document}